\documentclass[aps,preprint,showpacs]{revtex4}
\usepackage[T1]{fontenc}
\usepackage[latin1]{inputenc}
\usepackage{amsmath}
\usepackage{graphics}
\usepackage{graphicx}
\usepackage{amssymb}
\usepackage{dcolumn}
\usepackage{bm}
\usepackage{amsfonts}

\begin{document}

\preprint{}
\title{Search for excited electrons through $\gamma\gamma$ scattering}
\author{A. Ozansoy}
\email{aozansoy@science.ankara.edu.tr}
\affiliation{Ankara University, Faculty of Sciences, Department of Physics, 06100, Tandogan, Ankara, Turkey}
\author{A.A. Billur}
\email{abillur@cumhuriyet.edu.tr}
\affiliation{Department of Physics, Cumhuriyet University, 58140, Sivas, Turkey}

\begin{abstract}
We study the potential of $\gamma \gamma$ option  of future high energy linear $e^{+}e^{-}$ colliders to search for excited electrons with spin-1/2. We calculate single production cross sections, give the angular distributions and $f-m^{}*$ contour plots for $\sqrt{s}=0.5$ TeV and $\sqrt{s}=3$ TeV both using the standard (tree level) couplings and anomal couplings. 
\end{abstract}

\maketitle


\section{Introduction}

The number of the Standard Model (SM) fermionic families and replication of them still remains its mystery. 
A natural explanation for the replication of Standard Model (SM) fermionic families is lepton and quark compositeness. 
Composite models predict that  known fermions are made of more fundamental constituents called preons and preons 
interact by more strong interactions. An unambiguous consequence of such an underlying substructure is the appearence of excited 
states. In composite models known quarks and leptons can be regarded as the ground state to a rich spectrum of excited states.  Excited leptons $(l^{*})$ can appear in the framework of composite models \cite{Terazawa77, Renard83, Kuhn84}. 
The phenomenology of excited leptons have been searched at DESY Hadron Electron Ring Acceleartor (HERA) and 
CERN Large Electron Positron (LEP)\cite{Hagiwara85,Boudjema93}, Fermilab Tevatron \cite{Acosta05} and Large Hadron Collider (LHC)\cite{Cakir03,Baur90} and next linear colliders \cite{Cakir04,Cakir08}. Also, exclusive excited leptons search in the $pp \rightarrow pl^{-}l^{+}p$ reaction at the LHC was studied in \cite{Inan10}.

Up to now, no signal had been observed for excited states but next high energy particle colliders 
will be able to broaden the excited leptonic states searches. The LHC is expected to enlighten some 
of the questions left open by the SM with its high center of mass energy and high luminosity in the 
next few years. For precision measurements, lepton colliders must be build to supplement the hadron 
colliders. Next $e^{+}e^{-}$ linear colliders namely International Linear Collider (ILC)\cite{ILC07} and Compact Linar Collider (CLIC)\cite{CLIC12} would be complementary for the TeV scale physics studies at the LHC. 
Future high energy $e^{+}e^{-}$ colliders could have $e \gamma$ and $\gamma \gamma$ modes by converting the orijinal $e^{+}$ or $e^{-}$ beam into photon beam through Compton backscattering mechanism. The center of mass energies and luminosities for photon colliders are approximately same as for the basic $e^{+}e^{-}$ colliders \cite{Ginzburg92}. An elaborate study about the $\gamma  \gamma$ option of the $e^{+}e^{-}$ colliders has been given in \cite{Ginzburg83, Ginzburg84}.
These modes introduces new production channels of excited leptons such as single and pair production of excited charged leptons in $\gamma \gamma$ collisions and resonant production of excited electrons in $e \gamma$ mode. While only the excited electron can be produced singly in $e \gamma$ mode, it is possible to produce all generations of charged excited leptons both
 singly and in pairs in the $\gamma \gamma$ mode \cite{Boudjema93}.

The mass limits for excited electrons from their single production are $m^{\ast}>272$ GeV \cite{Aaron08} from H1 experiment at HERA assuming $f=f^{'}=\lambda/m^{\ast}$, from their pair production are $m^{\ast}>103.2$ GeV \cite{Abbiendi02} from LEP OPAL assuming $f=f^{'}$, and from indirect searches LEP L3 experiment excluded the mass region of $m^{\ast} > 310$ GeV \cite{Achard02}.
Recently, the ATLAS Collaboration has set the excited electron mass limits in the $l^{\ast} \rightarrow l \gamma$ decay channel at $\sqrt{s}=7$ TeV with an integrated luminosit $L_{int}=2.05$ fb$^{-1}$ assuming $\Lambda = m^{\ast}$ as $m^{\ast}< 1.87$ TeV excluded at $95\%$ C.L. \cite{Aad12}.

This study is a continuation of previous searches on excited electrons with spin-1/2 \cite{Cakir04, Ozansoy10}. In this work, in Sec.II  we introduce the effective lagrangians describing the gauge interactions of excited leptons with spin-1/2 both for single and pair productions.  In Sec.III, we analyze the signal and backgrounds for the process $\gamma \gamma \rightarrow  e^{-*}e^{+}$. We summarize our results in Sec.IV. In our calculations we use the simmulation programme COMPHEP-4.5.1 \cite{Pukhov99}.

\section{Excited Leptons}

Excited leptons can be classified by $SU(2)\times U(1)$ quantum numbers. A spin-1/2 excited lepton is the lowest radial and orbital excitation that has magnetic transition type interactions with the ordinary leptons. Excited leptons with higher spins are studied in \cite{Lopes80,Eboli96,Cakir08}. A $SU(2)\times U(1)$  invariant effective Lagrangian that describes the interaction between an ordinary lepton, a gauge boson and an excited lepton is given

\begin{equation}
L_{ll^{\ast}V}=\frac{1}{2\Lambda}\bar{l_R}^{\ast}\sigma^{\mu\nu}\left[fg\frac{\vec{\tau}}{2}\cdot\vec{W}_{\mu\nu}+f^{'}g^{'}\frac{Y}{2}B_{\mu\nu}\right]l_{L}+h.c.\end{equation}%

where $l, l^{\ast}$ denotes the ordinary and excited lepton, respectively, $V=W,Z,\gamma$, $\Lambda$ is the compositeness scale,$\sigma^{\mu\nu}=\frac{i}{2}[\gamma^{\mu}\gamma^{\nu}-\gamma^{\nu}\gamma^{\mu}]$ with $\gamma^{\mu}$ being the Dirac matrices, $W_{\mu\nu}$ and $B_{\mu\nu}$ are the field strength tensors, $\vec{\tau}$ deneotes the Pauli matrices, $Y$ is hypercharge, $g$ and $g^{'}$are the SM gauge coupling of $SU(2)$ and $U(1)$, respectively, and $f$ and $f^{'}$ are the new couplings for the corresponding gauge couplings that are related to the compositeness dynamics. In order to have $ll^{\ast} V$ interaction is $SU(2)_L \times U(1)_Y$ gauge invariant, it's characteristic must be tensorial. In other words there should be magnetic transition type interactions between $l$ and $l^{\ast}$ \cite{Hagiwara85,Boudjema93,Kuhn84}. Due to the chirality conservation, an excited lepton should not couple to both left- and right-handed ordinary leptons to prevent SM leptons to have a large anomalous magnetic moment \cite{Brodsky80}. The effective lagrangian gives the following interaction vertex of excited lepton.

\begin{equation}
\Gamma_{\alpha}^{l l^{\ast} V}=\frac{g_e}{2\Lambda}f_{V}q^{\beta}\sigma_{\alpha \beta}(1-\gamma_{5})\end{equation}

Here, $q$ is the vector boson's four-momentum and $g_{e}=\sqrt{4\pi\alpha}$ and $f_V$' s are the new weak and electromagnetic couplings. In terms of the third component of the weak isospin $(I_{3L})$, electric charge $e_f$, weak mixing angles's sine $(s_W)$ and cosine $(c_W)$ and new couplings $f$ and $f^{'}$; $f_{\gamma}, f_W$ and $f_Z$ are

\begin{equation}
f_{\gamma}=e_ff^{'}+I_3L(f-f^{'}), f_W=\frac{1}{\sqrt{2}s_W}, f_Z=\frac{4I_{3L}(c_W^2f+s_W^2f^{'}-4e_fs_W^2f^{'})}{4s_Wc_W}
\end{equation}

The interaction lagrangian that describes the two excited lepton and a gauge boson is vector-like, this lagrangian is given,

\begin{equation}
L_{l^{\ast}l^{\ast}V}=\bar{l_L}^{\ast}
\gamma^{\mu}\left[g\frac{\vec{\tau}}{2}\cdot\vec{W_{\mu}}+g^{'}\frac{Y}{2}B_{\mu}\right]l_{L}^{\ast}+h.c.\end{equation}

Since excited leptons have high masses and their composite nature, the interaction vertices derived from this lagrangian should include form factors and also anomalous magnetic moments. The most general form of the vertices  given in \cite{Boudjema93}

\begin{equation}
\Gamma_{\mu}^{l^{\ast}l^{\ast}V}=ig_{e}\left[f_{1}^{V}\gamma_{\mu}+\frac{i}{2m^{\ast}}f^{V}_{2}\sigma_{\mu\nu}q^{\nu}+ f^{V}_{3}\gamma_{\mu}\gamma_{5}+\frac{i}{2m^{\ast}}f^{V}_{4}\sigma_{\mu\nu}\gamma_{5}q^{\nu}\right]
\end{equation}

Here, $f_{1}$ refers to the charge form factor, $f_{2}$ to the anamolous magnetic moment, $f_{3}$ is the parity violating term (due to the $U(1)_{em}$ gauge invariance this term vanishes for real photons) and the CP violating term $f_{4}$ represents the electric dipole moment for photons. In the case where the compositeness scale is too large, the form factors $f_i$ degrade to the point-like tree-level (standard) form factors. In this limit, $f_4$ is absent, $f_{1} = e_f$ and $f_{2} = 0$. In the presence of form factors (anomal couplings), $f_{1} = e_{f}(1+s/\Lambda^{2})$ and $f_{2} = m^{\ast 2}/\Lambda^2$.

An excited lepton performs two-body decay processes resulting the decay into a gauge boson and a SM lepton.
The possible decay modes are; radiative decays $l^{\ast} \rightarrow l\gamma$, neutral weak current decays $l^{\ast} \rightarrow\ lZ$ and charged weak current decays $l^{\ast} \rightarrow \nu W$. Neglecting the ordinary
lepton masses, decay width of excited lepton;

\begin{equation}
\Gamma(l^{\ast} \rightarrow lV)=\frac{\alpha m^{\ast^{3}}}{4\Lambda^{2}}f_{V}^{2}(1-\frac{m_{V}^{2}}{m^{\ast^{2}}})^{2}(1+\frac{m_{V}^{2}}{2m^{\ast^{2}}})\end{equation}

For an excited electron, new couplings are given $f_{\gamma}=-(f+f^{'})/2$,
$f_{Z}=(-f\cot\theta_{W}+f^{'}\tan\theta_{W})/2$, $f_{W}=f/(\sqrt{2}\sin\theta_{W})$. The branching ratios BR$(\%)$ depending on the excited spin-1/2 electron for $f=f^{'}=1$ and $f=-f^{'}=1$ are given in \cite{Cakir08}.

\section{Signal and Backgrounds}

\begin{figure}
\includegraphics{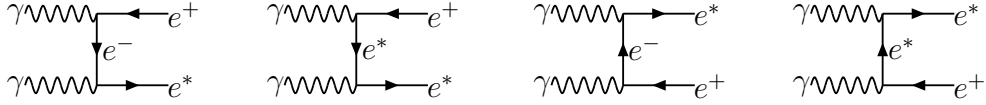}
\caption{Feynman diagrams for excited electron production via the process 
$\gamma \gamma \rightarrow e^{-\ast} e^{+}$.}
\label{fig1}
\end{figure}

Excited leptons have contact or gauge mediated interactions with the SM particles. Contact interactions could enlarge the discovery limits for future colliders \cite{Cakir03,Baur90}. Here we only concentrate on the gauge interactions of excited electrons. Excited electrons can be produced singly through $\gamma \gamma$ scattering. Feynmann diagrams for the process $\gamma \gamma \rightarrow e^{-*} e^{+}$ is given in Fig. \ref{fig1}.

\begin{figure}[ptbh]
{{}}\includegraphics[
height=6cm,
width=8cm
]{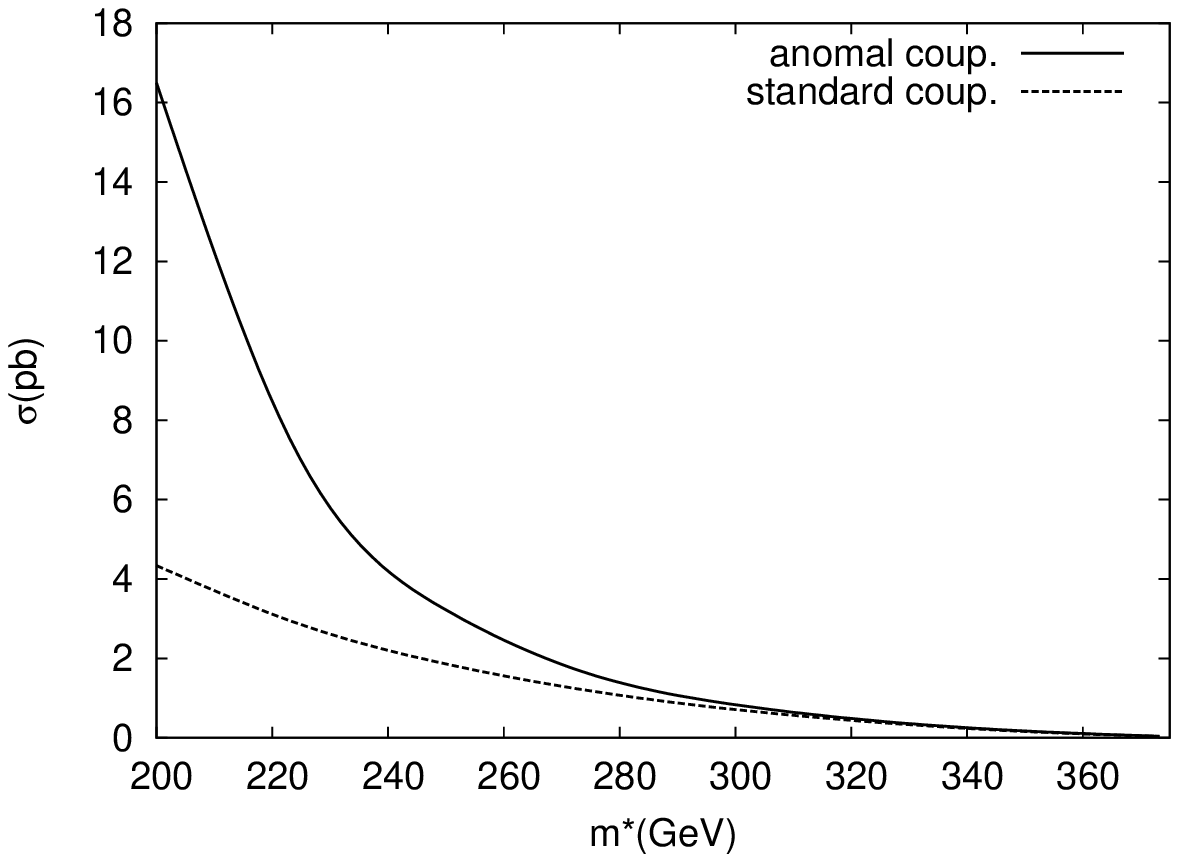}\includegraphics[
height=6cm,
width=8cm
]{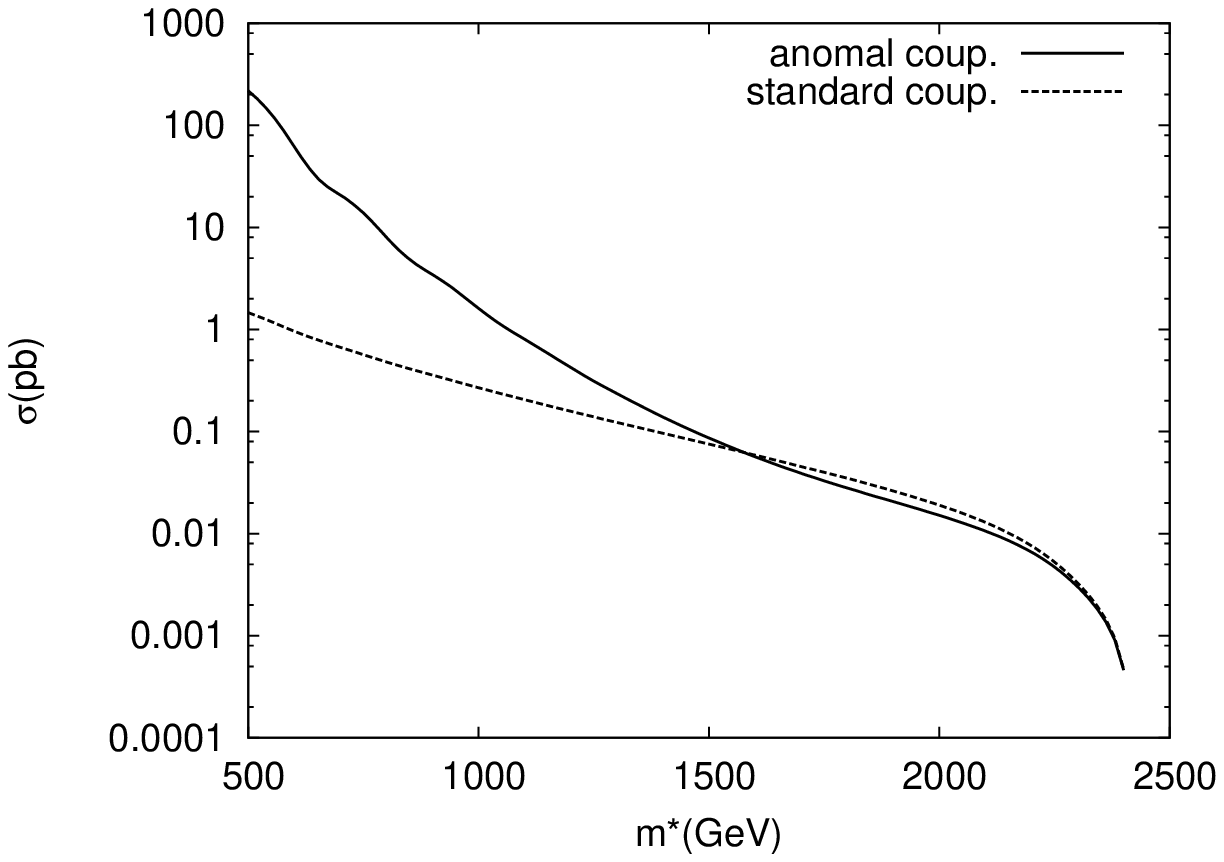}\caption{Total cross section as a function of the excited electron mass both for standard and anomal couplings 
at  $\protect\sqrt{s}=0.5$ TeV (left panel) and $\protect\sqrt{s}=3$ TeV(right panel).}%
\label{fig2}%
\end{figure}

The total cross section versus $m^{\ast}$ for the process $\gamma \gamma \rightarrow e^{-\ast}e^{+}$ both for standard couplings and anomal couplings is given in Fig. \ref{fig2}. Here we choose the compositeness scale dynamically $\Lambda = m^{\ast}$ and $f=f^{'}$ for precise measurements. For small excited electron masses, total cross sections for the anomal couplings distincly greater than the cross sections for the standard couplings. 

We choose the $e^{\ast}\rightarrow e\gamma$, electromagnetic decay mode of the excited electron for an easy detection. We apply the following acceptance cuts

\begin{align}
p_{T}^{e,\gamma }& >20\text{ GeV} \\
\left\vert \eta _{e,\gamma }\right\vert & <2.5 \\
\Delta R_{(e^+e^-),(e^{\pm}\gamma)} & >0.4 
\end{align}

where $p_{T}$ is the transverse momentum of the final state detectable particle,
$\eta $ denotes pseudorapidity, $\Delta R$ is the seperation of two of them. After applying these cuts we find the SM background cross section (for the $\gamma \gamma \rightarrow e^+e^- \gamma$ process) $\sigma_B = 2.23 \times 10^{-1}$ pb for $\sqrt{s}=0.5$ TeV and $\sigma_B = 5.11 \times 10^{-2}$ pb for $\sqrt{s}=3$ TeV.

\begin{figure}[ptbh]
{{}}\includegraphics[
height=6cm,
width=8cm
]{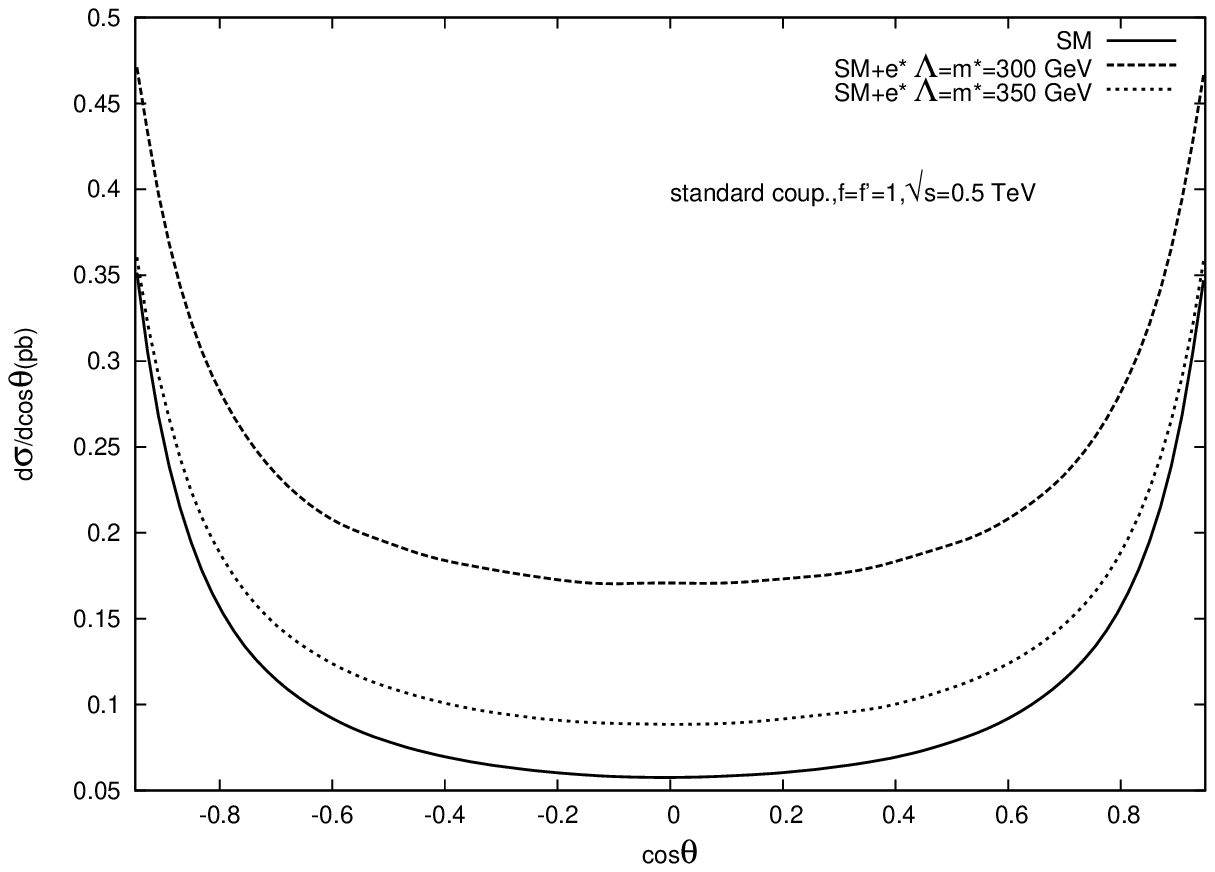}\includegraphics[
height=6cm,
width=8cm
]{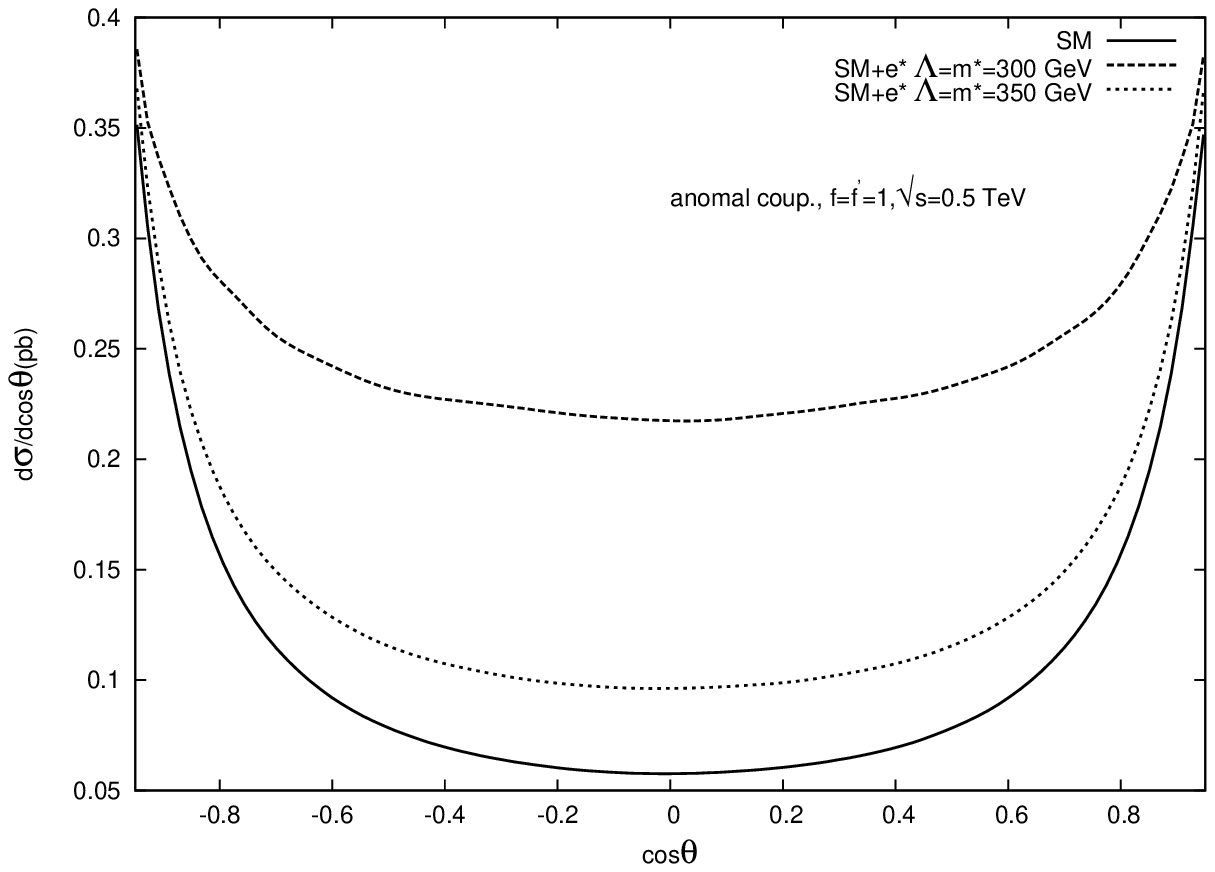}\caption{Angular distributions at  $\protect\sqrt{s}=0.5$ TeV for standard (left panel) and anomal (right panel) couplings.}%
\label{fig3}%
\end{figure}

\begin{figure}[ptbh]
\includegraphics[
height=6cm,
width=8cm
]{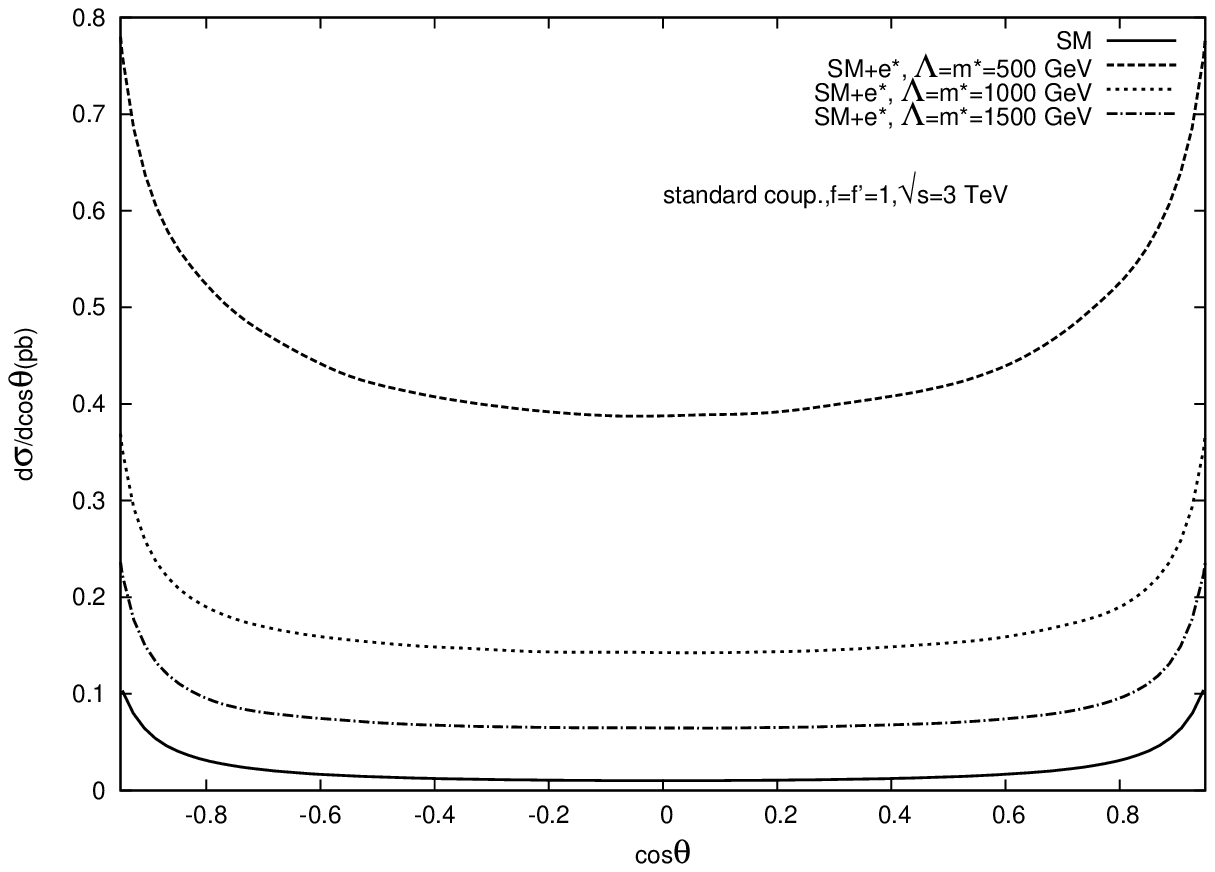}\includegraphics[
height=6cm,
width=8cm
]{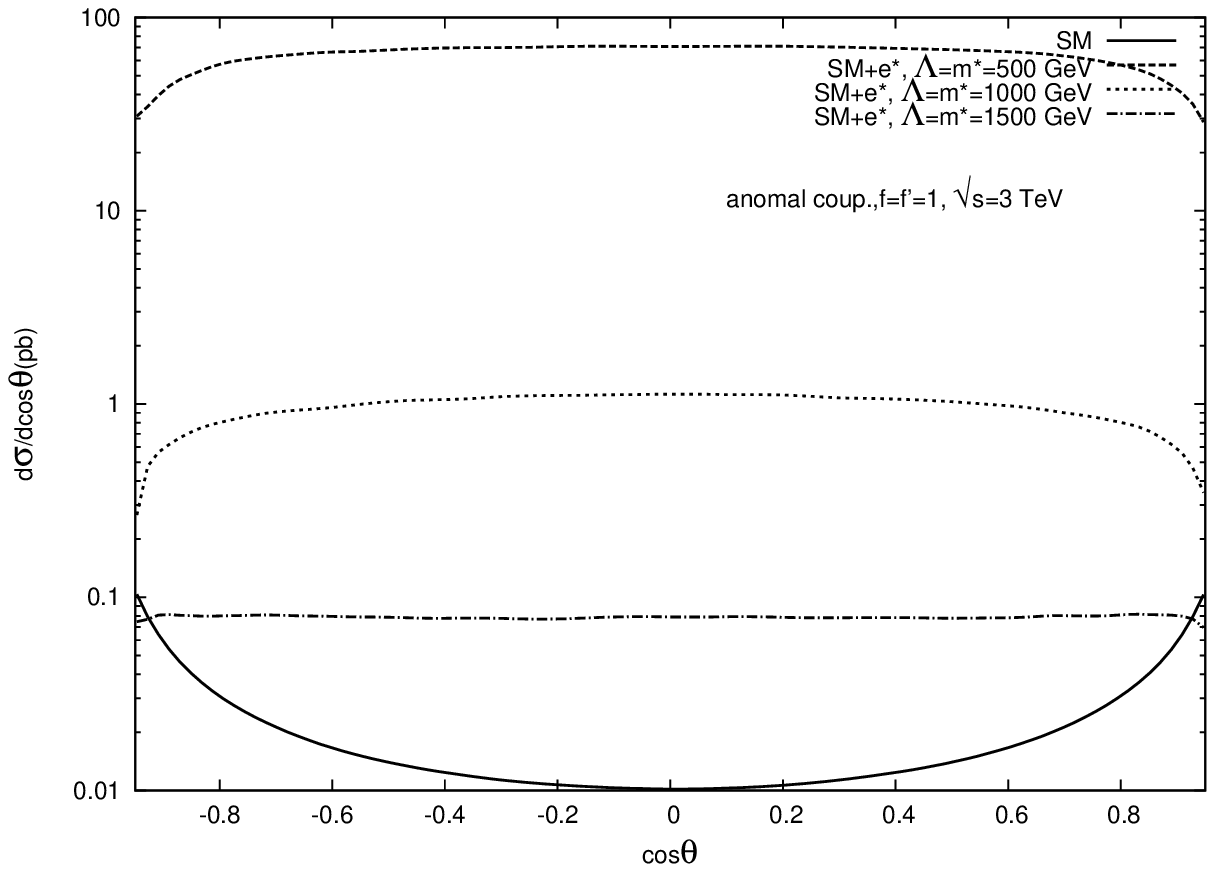}\caption{Angular distributions at  $\protect\sqrt{s}=3$ TeV for standard (left panel) and anomal (right panel) couplings .}%
\label{fig4}%
\end{figure}

In order to differantiate the excited electron signal from the SM background we plot the angular distributions for the process $\gamma \gamma \rightarrow e^{+}e^{- \gamma}$ for standard and anomal couplings at $\sqrt{s}=0.5$ TeV and $\sqrt{s}=3$
TeV in Fig. \ref{fig3} and Fig. \ref{fig4}, respectively. In these figures, it is pronounced that, excited electron signal is above the SM background. 

\begin{figure}
\includegraphics[
height=6cm,
width=8cm
]{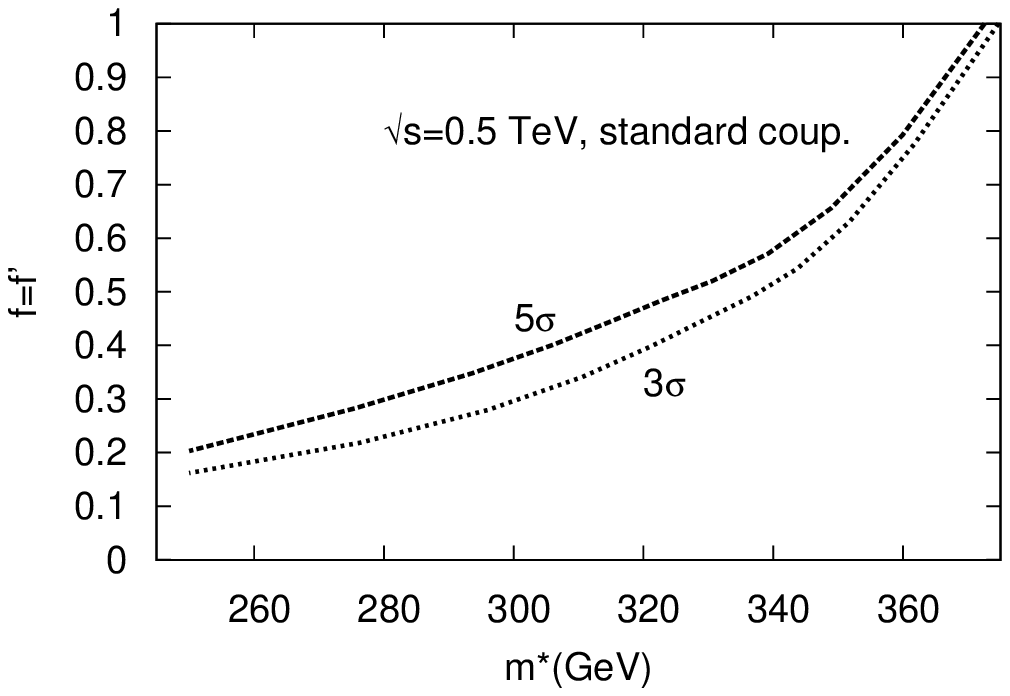}\includegraphics[
height=6cm,
width=8cm
]{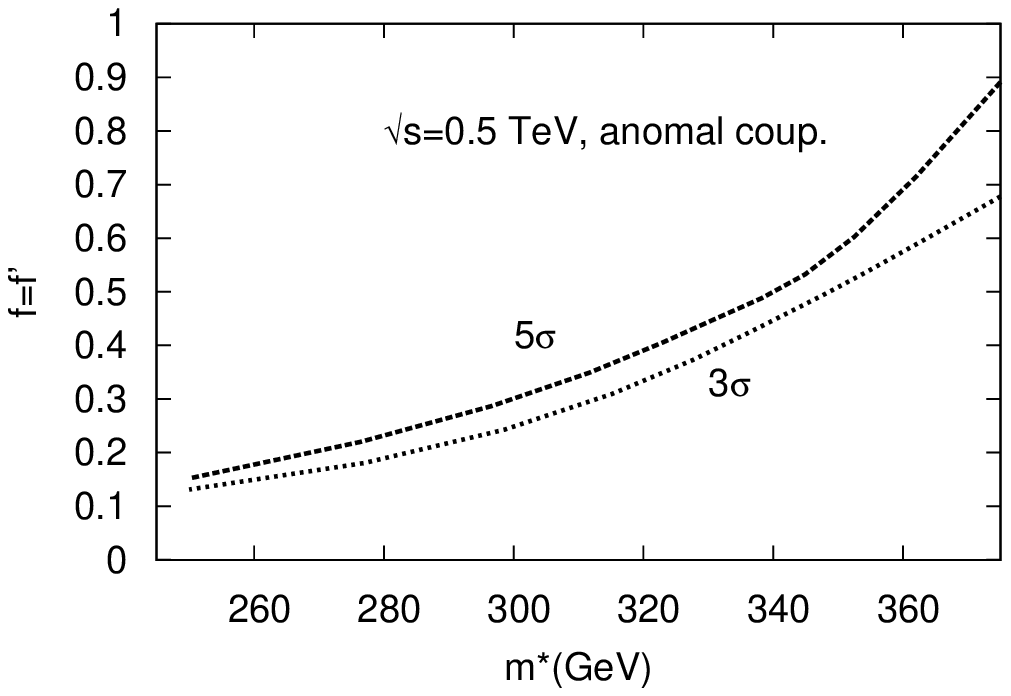}
\caption{Contour plots for  $\protect\sqrt{s}=0.5$ TeV for standard (left panel) and anomal (right panel) couplings. }
\label{fig5}
\end{figure}

\begin{figure}
\includegraphics[
height=6cm,
width=8cm
]{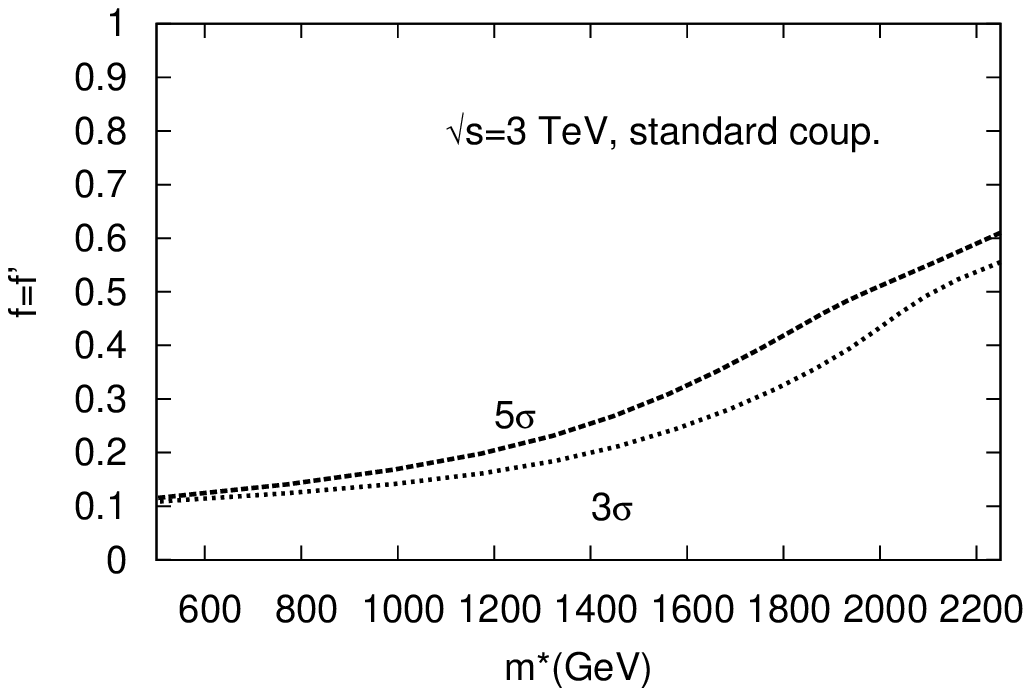}\includegraphics[
height=6cm,
width=8cm
]{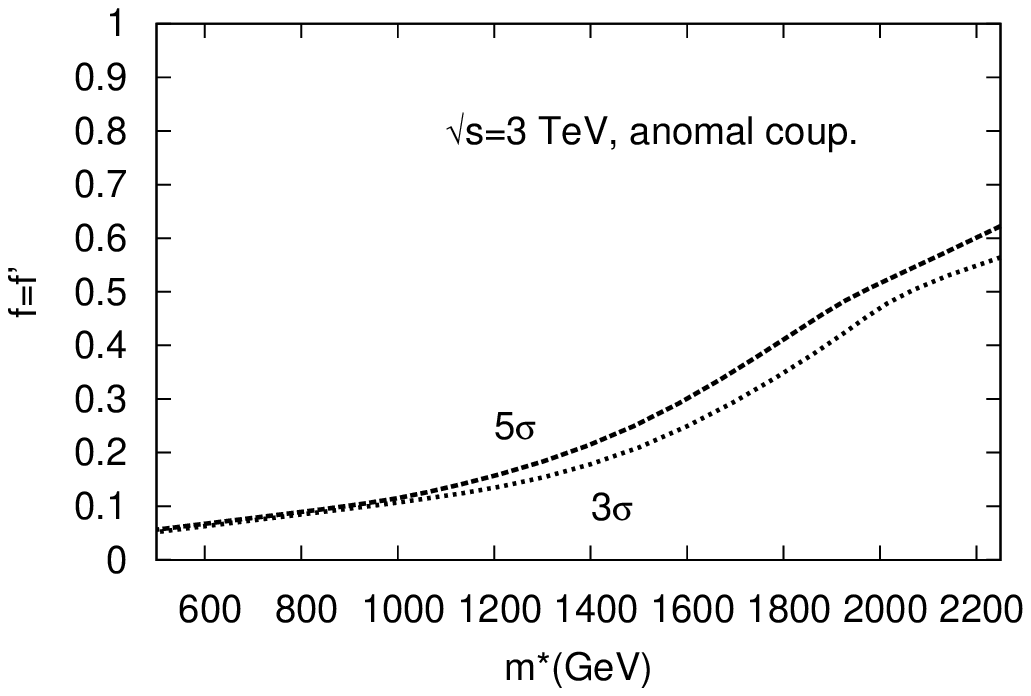}
\caption{Contour plots for $\protect\sqrt{s}=3$ TeV for standard (left panel) and anomal (right panel) couplings.}
\label{fig6}
\end{figure}

To analyze the potential of the $\gamma \gamma$ option of the high energy linear $e^{+}e^{-}$ colliders to probe the excited electrons, we define the statistical significance of $SS$ of the signal

\begin{equation}
SS=\frac{|\sigma_{S+B}-\sigma_{B}|}{\sqrt{\sigma_B}}\sqrt{L_{int}}
\end{equation}

where $L_{int}$ is the integrated luminosity of the $\gamma \gamma$ option of the linear collider. In the $f-m^{\ast}$ parameter space, we plot the contour plots for $3\sigma$ and $5\sigma$ deviations from the SM background only considering the acceptance cuts. We display our results for standard and anomal couplings for the center of mass energies of the $e^{+}e^{-}$ collider $\sqrt{s}=0.5$ TeV and $\sqrt{s}=3$ TeV in Fig. \ref{fig5} and Fig. \ref{fig6}, respectively. 

Concerning the criteria $SS>5$, with an integrated luminosity $L_{int} = 100$ fb$^{-1}$ and $\sqrt{s}=0.5$ TeV ILC can probe excited electrons of mass $m^{\ast}=280 $ GeV for standard interactions and $m^{\ast}=300$ GeV for anomal interactions for the couplings $f=f^{'}=0.3$. Taking into account the same criteria CLIC with an integrated luminosity $L_{int} = 200$ fb$^{-1}$ and $\sqrt{s}=3$ TeV can probe excited electrons of mass $m^{\ast}=1200 $ GeV for standard interactions and $m^{\ast}\approx 1300$ GeV for anomal interactions for the couplings $f=f^{'}=0.2$

\section{Conclusion}

$\gamma \gamma$ colliders are very natural complimentals to high energy linear $e^{+}e^{-}$ colliders. They are convenient to study the New Physics effects with energies and luminosities rather close to those in $e^{+}e^{-}$ colliders. In our study, we present the $\gamma \gamma$ option of the next linear colliders are promising for the observation of excited electrons with rather small couplings and high masses. We show that the limits on the excited electron mass $m^{\ast}$ for standard couplings and anomal couplings slightly differ for a given value of the new couplings $f=f^{'}$. The limits on the $m^{\ast}$ and $f, f^{'}$ stringently depend on the luminosity. A detailed study on effects restricts the $\gamma \gamma$ luminosity can be found in \cite{Telnov90}.

We only take into consideration the gauge interaction of spin-1/2 excited electrons with SM particles. It is also possible to do same analysis for excited electrons with spin-3/2. It is a superiority of the $\gamma \gamma$ collider that the single and double production of all generations of excited leptons possible through $\gamma \gamma$ collisions. Here we only studied the excited electrons; but excited leptons appear in three families. Our work can be extended for excited muon $\mu^{\ast}$ and excited tau $\tau^{\ast}$.

\begin{acknowledgements}
The work of A.O. was supported in part by the State Planning Organization (DPT) under Grant No. DPT-2006K-120470.
\end{acknowledgements}

\end{document}